\documentstyle[12pt]{article}
\begin{document}

\def\beq{\begin{equation}}
\def\eeq{\end{equation}}
\def\bea{\begin{eqnarray}}
\def\eea{\end{eqnarray}}
\def\ra{\rightarrow}
\def\D0{D\O~}
\def\CPbar{\hbox{{\rm CP}\hskip-1.80em{/}}}
\def\ETslash{\not{\hbox{\kern-4pt $E_T$}}}
\def\pbarp{ \bar{{\rm p}} {\rm p} }
\def\pp{ {\rm p} {\rm p} }
\def\ifb{${\rm fb}^{-1}$}
\def\del{\partial }
\def\ov{\overline}
\def\Tr{{\rm Tr}}
\def\pslash{\not{\hbox{\kern -1.5pt $p$}}}
\def\kslash{\not{\hbox{\kern -1.5pt $k$}}}
\def\aslash{\not{\hbox{\kern -1.5pt $a$}}}
\def\bslash{\not{\hbox{\kern -1.5pt $b$}}}
\def\Dslash{\not{\hbox{\kern -4pt $D$}}}
\def\wslash{\not{\hbox{\kern -4pt $\cal W$}}}
\def\zslash{\not{\hbox{\kern -4pt $\cal Z$}}}
\def\ttz{{\mbox {\,$t$-${t}$-$Z$}\,}}
\def\tta{{\mbox {\,$t$-${t}$-$A$}\,}}
\def\tbw{{\mbox {\,$W$-${t}$-$b$}\,}}
\def\tcZ{{\mbox {\,$t$-${c}$-$Z$}\,}}
\def\tuZ{{\mbox {\,$t$-${u}$-$Z$}\,}}
\def\tcg{{\mbox {\,$t$-${c}$-$g$}\,}}
\def\tsmdecay{$t \ra b \; W^+ \ra b \; \bar{\ell} \; {\nu_{\ell}} \;$}
%----------
\def\kln{\kappa_{L}^{NC}}
\def\krn{\kappa_{R}^{NC}}
\def\klc{\kappa_{L}^{CC}}
\def\krc{\kappa_{R}^{CC}}
\def\ttz{{\mbox {\,$t$-${t}$-$Z$}\,}}
\def\bbz{{\mbox {\,$b$-${b}$-$Z$}\,}}
\def\tta{{\mbox {\,$t$-${t}$-$A$}\,}}
\def\bba{{\mbox {\,$b$-${b}$-$A$}\,}}
\def\tbw{{\mbox {\,$t$-${b}$-$W$}\,}}
\def\tltlz{{\mbox {\,$t_L$-$\overline{t_L}$-$Z$}\,}}
\def\blblz{{\mbox {\,$b_L$-$\overline{b_L}$-$Z$}\,}}
\def\brbrz{{\mbox {\,$b_R$-$\overline{b_R}$-$Z$}\,}}
\def\tlblw{{\mbox {\,$t_L$-$\overline{b_L}$-$W$}\,}}
%-----------
\def\to{\rightarrow}
\def\Psibar{\bar{\Psi}}
\def\phibb{\phi b \bar{b}}
\def\hbb{h b \bar{b}}
\def\bbbb{b \bar{b} b \bar{b}}
\def\bbjj{b \bar{b} j j}
\def\Zbb{Z b \bar{b}}
\newcommand{\tanb}{\tan\beta}
%-------------
\setcounter{footnote}{1}
\renewcommand{\thefootnote}{\fnsymbol{footnote}}
%=========================Title Page=========================
\begin{titlepage}

{\small
\noindent
{hep-ph/9809536} \hfill {MSUHEP-80828} \\
{August 1998} \\
}

\vspace{2.0cm}
%\vspace{1.0cm}
 
\begin{center}
{\Large\bf Top Quark and Electroweak Symmetry Breaking Mechanism 
\footnote{
\baselineskip=12pt  
Talk given at the International Seminar ``Quarks-98'',
May 17-24, 1998, Suzdal, Russia.}
}
\end{center}

%\vspace*{1.2cm}
\vspace*{0.2cm}

\baselineskip=17pt
\centerline{\normalsize  
C.--P. Yuan}
 
\centerline{\normalsize\it
Department of Physics and Astronomy, Michigan State University }
\centerline{\normalsize\it
East Lansing, Michigan 48824 , USA }

\vspace{2cm}

\raggedbottom
\relax

\begin{abstract}
\noindent
After a brief comment on the role of top quark in models of electroweak 
symmetry breaking (EWSB), I shall discuss what we know about
top quark interaction and how to improve that knowledge.
Since bottom quark is the weak isospin partner of the top quark,
its interaction with a scalar boson may also distinguish models of
EWSB. We show that Tevatron can provide useful information through
the associated production of a scalar with a bottom quark pair.
\end{abstract}

%\vspace*{3.4cm}
%PACS numbers: 14.65.Ha, 12.39.Fe, 12.60.-i

\end{titlepage}
%===============end of Title Page=========================

%\vspace*{0.2cm}

%\pagestyle{plain}
%\setcounter{page}{1}

\renewcommand{\thefootnote}{\arabic{footnote}}
\setcounter{footnote}{0}
\baselineskip=18pt 

\section{Introduction}
\label{intro} 

Two of the outstanding mysteries in electroweak theory are: 
(i) the cause of the electroweak symmetry breaking (EWSB), which 
gives masses to the weak gauge bosons $W^\pm$ and $Z$, and
(ii) the origin of flavor symmetry breaking (FSB), which makes 
quarks and leptons have diverse masses.
In the standard model (SM) of particle physics, both symmetry 
breaking mechanisms
 are accommodated by including a fundamental weak doublet
of scalar (Higgs) boson. Because of the spontaneous symmetry breaking,
the Higgs boson develops a vacuum expectation value (VEV) $v$,  
so that
the weak gauge bosons gain their masses. Similarly, fermion gains
mass via Yukawa interaction with Higgs boson. 
However, SM provides no explanation (i.e., no explicit dynamics)
for the generation of mass. 
 
The models of symmetry breaking beyond the SM 
can be characterized into two 
classes. One is the weakly interacting model
 (with elementary Higgs bosons), such as 
supersymmetry (SUSY) theory \cite{susy}, 
another is the strongly interacting model (with composite Higgs bosons),
such as Topcolor model \cite{hill}.
In these models, top quark often plays a very special role in 
the EWSB and/or the FSB dynamics.
For example, in the supergravity model, the electroweak symmetry is
broken radiatively by corrections from top quark and top-squarks 
(superpartners of left- and right-handed top quark) in the self-energy 
of Higgs boson \cite{susyrad}. 
Because top quark is heavy ($\sim v/\sqrt{2}$),
the running mass of the Higgs boson field becomes negative around 
TeV energy scale where the spontaneous symmetry breaking occurs.
Thus, EWSB is driven by a heavy top quark.
In the Topcolor-assisted Technicolor (TCATC) model \cite{tcatc}, 
some unknown strong dynamics in the Techni-fermion sector 
induces electroweak symmetry breaking through the formation 
of the Techni-fermion condensation, which generates the masses
of $W^\pm$ and $Z$ gauge bosons.
What is the role of top quark? 
In the usual extended Technicolor (ETC) model, fermions gain mass by 
interacting with ETC gauge bosons and Techni-fermions.
It is a fine idea to generate masses for light fermions, but
in order to give a large mass to the top quark, the mass of the
ETC gauge boson has to be small. Unfortunately, in that case, 
the model would predict a large shift ($\Delta \rho$)
in the $\rho$-parameter, and is not tolerable by data.
This is a well known problem in ETC models. However, in the TCATC model,
top quark plays a very special role to solve this problem of FSB.
How does it work?
In the TCATC model, top quark (and bottom quark) experiences a new 
strong gauge (Topcolor) interaction so that a heavy top quark pair
can condensate and give a large (almost all) mass to top quark,
while the condensate contributes only a little to 
the breaking of the electroweak symmetry.
By this, it solves the $\Delta \rho$ problem and generates a more 
``natural'' ETC model to describe the fermion mass spectrum.
Obviously, the above ideas for either class of models will not work if
the mass of the top quark were not large enough.
In conclusion, it seems to be reasonable that a heavy top quark 
can play an important role in the electroweak symmetry breaking
and/or flavor symmetry breaking dynamics.

With the discovery of the top quark by the CDF and \D0
collaborations, it has become natural to
consider its properties, such as its couplings to the other particles.
By now, all the experimental data show excellent agreement with 
SM prediction. Does that imply new physics is not allowed and top quark
interaction is determined? No, that is not the case. 
The present data does not exclude possible new physics.
In the next section, I will discuss what data is telling us about 
the interaction of top quark.

\section{Constraints From Low Energy Data}
%\section{Constraints to Top Quark Interactions From Low Energy Data}

If we do not assume a SM top quark, do we know anything about the
interactions of top from low energy data, such as the precision
$Z$-pole data and bottom physics?
Will there be any surprise from future collider data, such as the
Fermilab upgraded Tevatron, CERN Large Hadron Collider (LHC), future
Linear Collider (LC)? If yes, how to look for them?
In this section, I will address the first question, and defer the others
to the next section.

We can perform the study in a model-independent way by constructing 
an effective low energy theory that is consistent with the symmetry
breaking pattern of the SM, i.e., 
the gauge symmetry $SU(3)_C \times SU(2)_L \times U(1)_Y$ is 
spontaneously broken down to $SU(3)_C \times U(1)_{em}$. 
A well established technique to construct a complete set of operators
that respects the symmetries in expansion of energy is to build
an electroweak chiral Lagrangian (EWCL), in which the 
$SU(2)_L \times U(1)_Y$ symmetry is non-linearly realized \cite{chiral}.
To simplify our study, we shall assume that the effect of 
heavy new physics is
to only modify the couplings of $W$-$t$-$b$ and $Z$-$t$-$t$
without introducing any non-SM light field.
In that case,
the most general gauge invariant chiral Lagrangian \cite{larios}, 
that includes the electroweak 
couplings of the top quark up to dimension four, 
contains terms such as
$\frac{1}{\sqrt{2}}
\left (1+\kappa_{L}^{\rm {CC}}\right ) \ov {{t}_{L}}
\gamma^{\mu} b_{L}
{{\cal W}_{\mu}^+}$,
$\frac{1}{\sqrt{2}}\kappa_{R}^{\rm {CC}}
\overline{{t}_{R}}\gamma^{\mu} b_{R}
{{\cal W}_{\mu}^+}$,
$\frac{1}{6} \left (3- 4 s_w^2 +
3 \kappa_{L}^{\rm {NC}}\right)
\overline{t_{L}}\gamma^{\mu} t_{L}{{\cal Z}_{\mu}}$,
and
$\frac{1}{6} \left ( -4s_w^2+
3 \kappa_{R}^{\rm {NC}}\right ) \ov {{t}_{R}}
\gamma^{\mu} t_{R}{{\cal Z}_{\mu}}$, where
$\kappa$'s parameterize possible new physics.
(Here, we do not include possible flavor-changing neutral current 
(FCNC) couplings, e.g. $t$-$c$-$Z$,
or dimension five operators \cite{larios}.)
In general, the charged current coefficients can be complex with the
imaginary part introducing a CP odd interaction, and the neutral current
coefficients are real so that the effective Lagrangian is hermitian.
In the unitary gauge, the {\it composite} fields 
${\cal W}_{\mu}^\pm$, ${{\cal Z}}_{\mu}$, and
${{\cal A}}_{\mu}$ are reduced to 
$-gW_{\mu}^\pm$, $-\frac{g}{c_w} Z_{\mu}$, 
and $\frac{e}{s_w^2}A_{\mu}$, respectively,
where $e=g s_w=g^{\hspace{.5mm}\prime} c_w$ and $s_w=\sin \theta_w$,
etc.

Top quark can only contribute to low energy observables through
loop corrections. To concentrate on the interactions of top quark and
EWSB sector, we shall only include those non-standard contributions 
(from the $\kappa$'s) of the order
${{m^2_t}\over {16 \pi^2 v^2}} \; \ln {{\Lambda^2}\over {m^2_t}}$, where
$\Lambda$ is a physical cutoff scale below which  
the effective Lagrangian is valid. 
(Here, $\Lambda$ is taken to be $4\pi v \sim 3$\,TeV.)

Since any contributions from the
right-handed charged current coupling $\krc$ are proportional to the
bottom quark's mass $m_b$ (which is much smaller than $m_t$), 
we can only obtain useful bounds for $\kln$, $\krn$ and $\klc$
from the $Z$-pole data at the LEP and the SLC, up to
one loop level.
However, $\krc$ can be studied independently by
using the CLEO measurement of the branching ratio ($BR$) of
 $b\ra s\gamma$, in which $\krc$ becomes the significant anomalous
coupling.  From the theoretical prediction \cite{fuj} and 
the CLEO measurement 
$1\times 10^{-4}<BR(b\ra s\gamma)<4.2\times 10^{-4}$ \cite{cleo},
it was found \cite{larios} that 
$-0.037 \; < \; \krc \; < 0.0015 \;$.
Hence, $\krc$ is strongly constrained for the case that there is no
new light field contributing to the $b \ra s \gamma$ data.
With these observations, we study how $\kln$, $\krn$ and
$\klc$ can be constrained by LEP/SLC data, which,
under a few general
assumptions, can be parameterized by 4-independent parameters:
$\epsilon_1$, $\epsilon_2$, $\epsilon_3$, and $\epsilon_b$ \cite{epsilon}.
Namely, all the leading 
contributions of the non-standard couplings $\kappa$'s 
 are contained in the oblique corrections, i.e., the
vacuum polarization functions of the gauge bosons, and the non-oblique
corrections to the vertex \bbz. The non-standard 
contributions to the $\epsilon$ parameters
are:
$\delta\epsilon_1=\frac{G_F}{2\sqrt{2}{\pi}^2}3{m_t^2}
 (-\kln+\krn+\klc)\ln{\frac{{\Lambda}^2}{m_t^2}}\,\,$
and 
$\delta\epsilon_b=\frac{G_F}{2\sqrt{2}{\pi}^2}{m_t^2}
\left ( -\frac{1}{4}\krn+\kln \right ) \ln{\frac{{\Lambda}^2}{m_t^2}}
\,\,$.
It is interesting to note that $\klc$ does not contribute to 
$\epsilon_b$ up to this order ($m_t^2\ln {\Lambda}^{2}$).
Given the above results we can then use the experimental values
of the $\epsilon$'s to constrain the theoretical 
predictions. We find that
 precision data allows for all three non-standard
couplings to be different from zero.  There is a three
dimensional boundary region for these $\kappa$'s.
The only coefficient that is
constrained at the 95\% confidence level (C.L.)
is $\kln$ which can only vary between $-0.35$ and
$0.35$.
  The other two can vary through the
whole range ($-1.0$ to $1.0$) although in a correlated manner \cite{larios}.
(For instance, $\epsilon_b$ data implies $\krn \sim 4 \kln$ for any
$\klc$.)
Furthermore, LEP/SLC data imposes $\klc \, \sim \, - \krn$
if $\kln$ is close to zero.  This 
conclusion holds for $m_t$ ranging from
$160$ GeV to $180$ GeV.  
Hence, the precision low energy
data does not exclude the possibility 
of having anomalous top quark interactions
with the gauge bosons. 

Different models for the electroweak symmetry
breaking sector can induce different relations among the $\kappa$'s.
These relations can in turn be used to discriminate between models
by comparing their predictions with experimental data.
To illustrate this point, consider a model \cite{ehab} with
$\klc = {1 \over 2} \kln$ (due to an approximate custodial symmetry
which ensures that $\rho$-parameter is close to one)
and another model~\cite{pczh} with $\klc=0$. 
The allowed range predicted 
by these two models lies 
along the line $\kln =2\krn$ and $\kln=\krn$
(with,$-0.1 < \kln < 0.15$), respectively.
If we imagine that any prescribed dependence between the 
couplings corresponds to a symmetry-breaking scenario, then, given the 
present status of low energy data, it may be
possible to distinguish the above two 
scenarios if $\kln$, $\krn$ and 
$\klc$ are larger than 10\%.
One such model that predicts $\klc \sim 10\%$ can be found 
in Ref. \cite{kuang}. 
 
In conclusion, we shown that new physics possibility 
(e.g., with non-zero $\kappa$'s) is allowed by the 
current low energy data. Only direct measurement 
(not through loop effect) on 
$\kappa$'s can conclusively test the interaction of
top quark with gauge bosons.
That means we have to study the direct production of the top quark
at high energy colliders.

\section{Direct Measurement of Top}
%\section{Direct Measurement of Top Quark Interactions}

While production of $t \bar{t}$ pairs
provides an excellent opportunity to probe the top's
QCD properties,
in order to carefully measure the top's electroweak interactions
it is also useful to consider single top production, in
addition to studying the decay of the top quark in $t \bar t$
events.

Single top production at a hadron collider occurs dominantly
through three sub-processes \cite{tim}. The $W^*$ mode of production
occurs when a quark and an
anti-quark fuse into a virtual $W$ boson, which then splits
into a $t$ and $\bar b$ quark. The $W$-gluon fusion mode occurs
when a $b$ quark fuses with a $W^{+}$ boson, producing a
top quark. The $t W^-$ mode occurs when a $b$ quark radiates a $W^-$.
The three single top production processes contain the \tbw
vertex of the SM, and thus are sensitive to 
the Cabibbo-Kobayashi-Maskawa (CKM) parameter $V_{tb}$
in the SM and to any possible
modification of this vertex from physics beyond the SM
(e.g, that generating a non-zero $\kappa^{CC}_{L,R}$).
The $t W^-$ process is important at the LHC,
but is highly suppressed at the Tevatron because
of the massive $W$ and $t$ particles in the final state.  
In Ref. \cite{tim}, a detailed analysis was carried out for the 
other two production processes at the Tevatron Run II energy
(a $\pbarp$ collider with $\sqrt{S}=2$\,TeV),
up to next-to-leading order (NLO) in QCD interaction.
The predicted theoretical values
and uncertainties of $\sigma_{W^*}$ and $\sigma_{Wg}$
(including single-$t$ and single-$\bar t$ rates) are:
$\sigma_{W^*} = 0.84 \; {\rm pb} \pm 15.5\%$ 
and 
$\sigma_{Wg}  = 2.35 \; {\rm pb} \pm 10.2\%$,
for a 175\,GeV top quark.
The theoretical uncertainties from scale, 
parton distribution functions (PDF),
and uncertainty in $m_t$ (assuming $m_t = 175 \pm 2$ GeV)
are found to be $\pm 5\%$ ($\pm 4\%$), $\pm 2\%$ ($\pm 3\%$),
and $\pm 6\%$ ($\pm 3\%$), for
$\sigma_{W^*}$ ($\sigma_{Wg}$).
Hence, 
the total uncertainties obtained from adding these uncertainties in
quadrature, linearly,
and the result from the envelope method
are $\pm 8\%$ ($\pm 6\%$), $\pm 13\%$ ($\pm 10\%$), and
$\pm 15.5\%$ ($\pm 10.2\%$), for $\sigma_{W^*}$ ($\sigma_{Wg}$).
To determine how well $\sigma_{W^*}$ and $\sigma_{Wg}$ can be measured
{\it experimentally}, experimental systematic uncertainties 
must be included.
Assuming a semi-leptonic top decay into
an electron or muon, and the detection efficiencies of
$9\%$ for the $W^*$ process and $33\%$ for the $W$-gluon
fusion process obtained from leading order (LO) studies,
the projected statistical uncertainties for a 2 \ifb (10 \ifb) of
integrated luminosity are
$\pm 17\%$ ($\pm 8\%$) and $\pm 5\%$ ($\pm 2\%$) for
measuring  $\sigma_{W^*}$ and $\sigma_{Wg}$, respectively.
The corresponding total uncertainties (including theoretical and 
statistical uncertainties) are 
$\pm 23\%$ ($\pm 17\%$) and $\pm 11\%$ ($\pm 10\%$). 
In the SM, the square-root of the single-top cross section 
is proportional to $|V_{tb}|$.
Hence, the expected uncertainty in measuring $|V_{tb}|$ is 
about $\pm 10\%$ and $\pm 5\%$ from the $W^*$ and 
$W$-gluon fusion data, assuming $|V_{tb}|$ is close to 1.
Furthermore, the relevant
background rates (52 and 350 events, for a 2 \ifb of luminosity)
after the kinematic cuts are about the same as the 
signal rates (34 $W^*$ and 345 $W$-gluon fusion events).
It should be emphasized that the use of the LO backgrounds
and efficiencies is an approximation, however we
expect these estimates to correspond rather well to the true NLO
results.

We note that the $W$-gluon fusion mode,
within the SM, provides a way to directly measure
the partial width of the top quark, $\Gamma(t \ra W^+ b)$,
through the effective-$W$ approximation.
Under this approximation, $\sigma_{Wg}$ can
be related to the width $\Gamma(t \ra W^+ b)$ by
the equation \cite{tim}
\bea
\sigma_{Wg} \simeq \sum_{\lambda = 0, +, -} \;
\int dx_1 \; dx_2 \; f_{\lambda}(x_1) \; b(x_2)
\left[ \frac{16 \pi^2 m_t^2}{\hat s (m_t^2 - M_W^2)} \right]
\Gamma(t \ra W^+_{\lambda} b) \; ,
\nonumber \;
\eea
where $x_1 x_2 = \hat s / S$,
$f_\lambda(x_1)$ is the distribution function for
$W$ bosons within the proton carrying momentum fraction
$x_1$, $b(x_2)$ is the $b$ quark PDF, and
$\lambda$ is the polarization of the $W$ boson.
Thus, if one has an experimental measurement of the $W$-gluon fusion rate,
it can be combined with the known effective $W$ distribution functions
to extract the partial width.
(We note that there is no similar relation between
$\sigma_{W^*}$ and $\Gamma(t \ra W^+ b)$.)
This method relies on the fact that within the SM
there are no FCNC
interactions, and the CKM elements $V_{ts}$ and
$V_{td}$ are very small; thus the
$t$-channel single top production involves fusion of
only the $b$ parton with a $W^+$ boson\footnote{Assuming
a top mass of $m_t = 175$ GeV, including the non-zero
contributions from $V_{td} = 0.009$ and $V_{ts} = 0.04$
increases the $W$-gluon fusion cross section
by less than $0.5\%$.}.
Once this partial width has been extracted from a measurement
of $\sigma_{Wg}$, it can be combined with a measurement of the
branching ratio ($BR$) of $t \ra W^+ b$ 
(obtained from examining top decays within
$t \bar t$ production)
to get the top quark's full width ($\Gamma(t \ra X)$, where
$X$ is anything) via the relation
$\Gamma( t \ra X) = {\Gamma(t \ra W^+ b)} / 
{BR( t \ra W^+ b)} \;$.

As also noted in Ref. \cite{tim} that because of the
different sensitivities to different types of 
new physics effects of the two production modes, it is
useful to consider the experimental data in the
$\sigma_{W^*}$-$\sigma_{Wg}$ plane.  
For example,
$\sigma_{W^*}$ is sensitive to the presence of 
a new heavy resonance contributing through s-channel diagram,
while $\sigma_{Wg}$ is sensitive to top quark FCNC interactions
contributing through t-channel diagram.
Comparison of the predictions of explicit models with the experimental 
point on this plane could be used to rule out or constrain these
models.  Furthermore, the $t W^-$ mode of single
top production is insensitive to the types of new physics
mentioned above, and thus could provide a safe
way to measure $V_{tb}$, provided enough statistics or
a carefully tuned search strategy compensates for its low
cross section. 

In conclusion, it is important to study single top
production at the Tevatron, in both the $W^*$ and
$W$-gluon fusion modes separately, as these two modes
provide complimentary information about the top quark.
Single top production provides an excellent opportunity
to directly measure
$V_{tb}$, and to search for possible signs of the new physics
associated with the top quark.  
Besides all the potential physics discussed above,
the Tevatron, as a $p \bar{p}$ collider, is unique for being able 
to test CP violation by measuring the production rates 
of single-top events. A nonvanishing asymmetry in 
the inclusive production  rates of the single-$t$ events and the
single-$\bar t$ events signals CP violation \cite{tcp}.  
Thus it can be used to constrain or detect this type of new physics.

While Tevatron can provide sensitive test to the coupling
of $t$-$b$-$W$, it cannot say very much about 
the coupling of $Z$-$t$-$t$, which has to be studied at
the LHC via the production of $Zt {\bar t}$, or 
at the LC through s-channel $Z$-diagram contribution.
It is also possible that the same dynamics that modifies the
three-point vertices of top quark also modifies its
four-point vertices. This was discussed in Ref. \cite{larios}.

\section{From Bottom To Top}
%\section{Bottom Quark, as the Weak-Isospin Partner of Top}

As argued earlier, top quark may play a special role in the 
mechanism of EWSB and/or FSB. One of such ideas is that
some new strong dynamics may involve a composite Higgs
sector to generate the EWSB and
to provide a dynamical origin for the top quark 
mass generation ({\it e.g.,} 
the top-condensate/top-color models \cite{hill}).
Another idea is realized in the supersymmetric
theories in which the EWSB is driven radiatively by the large top quark
Yukawa coupling with some fundamental Higgs bosons \cite{susy}.

Since the third family $b$ quark, as the weak isospin
partner of the top, can  have large 
Yukawa coupling with the Higgs 
scalar(s) in both composite and supersymmetric models, 
it was proposed in Ref. \cite{hjhe} 
to use the $b$ quark as a probe of possible non-standard
dynamics in Higgs and top sectors.
Because of the light $b$ mass 
relative to that of the top,
the production of Higgs boson associated with $b$ quarks
($\pbarp, \pp \to \phibb \to \bbbb$) may be experimentally
accessible at the Tevatron
and the LHC, even though the large
top mass could render associated Higgs production
with top quarks ($\pbarp, \pp \to \phi t \bar{t}$) infeasible.
This makes it possible for the Tevatron
and the LHC to test various models in which the $b$-quark
Yukawa coupling is naturally enhanced relative to the SM prediction.
Using the complete tree level results of the signal 
and background rates (with an estimated QCD
$k$-factor of 2), we derive the exclusion contour for the enhancement
factor (in the coupling of $\phibb$ relative to that of the SM)
versus the Higgs mass $m_\phi$ at the $95\%$~C.L., assuming 
a signal is not found.
We apply these results to analyze the constraints on the parameter
space of both the composite models and 
the MSSM (in the large $\tanb$ region). 
For the composite Higgs scenario, 
we first consider the two-Higgs-doublet extension (2HDE) of 
top-condensate model~\cite{tt-2HDM} and then analyze the topcolor
model, where the $b$ quark Yukawa couplings are naturally large 
(about the same as the top quark Yukawa coupling, which is around 1)
due to the infrared quasi-fixed-point structure and the particular 
boundary conditions for $(y_b,y_t)$ at the compositeness scale. 
The Tevatron Run~II with a 2~fb$^{-1}$ of
luminosity can exclude the
entire parameter space of the simplest 2HDE of top-condensate model,
if a signal is not found.
For the topcolor model, 
the Tevatron Run~II is able to detect the composite Higgs
$h_b$ or $A_b$ up to $\sim 400$~GeV and the LHC can extend the
mass range up to $\sim 1$~TeV.

To confirm the MSSM, it is necessary to 
detect all the predicted neutral 
Higgs bosons $h,~H,~A$ and the charged scalars $H^\pm$.
From LEP~II, depending on the choice 
of the MSSM soft-breaking parameters,
the current $95\%$~C.L. bounds on the masses of the MSSM
Higgs bosons are about 70~GeV for both the $CP$-even scalar $h$ and
the $CP$-odd scalar $A$. It can be improved at LEP~II
with higher luminosity and maximal energy, but the bounds on the
Higgs masses will not be much larger than $\sim m_Z$ for an   
arbitrary $\tanb$ value. The $Wh$ and $WH$ associated production at
the Tevatron can further improve these bounds, if a signal
is not observed. At the LHC, a large portion of parameter space can be 
tested via $pp\to t\bar t +h(\to \gamma \gamma )+X$, and
$pp\to h(\to ZZ^\ast )+X$, etc. 
A future high energy $e^+e^-$ collider will fully test
the MSSM Higgs sector through the
reactions $ e^+ e^- \to Z+h(H),\, A+h(H), \, H^+ H^-$, etc. 
We concluded \cite{hjhe}
that studying the $\phibb$
channel at hadron colliders can further improve our knowledge on MSSM.
The exclusion contours on the $m_A$-$\tanb$ plane of the MSSM
shows that Tevatron and LHC 
are sensitive to a large portion of the parameter space
via this mode. Therefore, it provides a complementary   
probe of the MSSM Higgs sector 
(including both the supergravity and the gauge-mediated SUSY
breaking models) 
in comparison with that from LEP~II.
Thus, it is expected
that experimental searches for this signature at the Tevatron
and the LHC will provide interesting
and important information about the mechanism of 
the electroweak symmetry
breaking and the fermion mass generation.
Fortunately, we shall have data very soon.

\section*{Acknowledgments }
%\indent \indent

I thank the organizers for the warm hospitality, and 
V. Ilyin and A. Pukhov for introducing me Russia, a beauty country 
with many friendly people.
I would also like to thank my collaborators:
C. Balazs, D. Carlson, J.L. Diaz-Cruz, H.-J. He, G. Kane, G. Ladinsky,
F. Larios, E. Malkawi, S. Mrenna,
and T. Tait, from them I have learned a great deal about
top quark physics.
This work  was supported in part by the NSF grant
No. PHY-9802564.

\end{document}